\documentclass[journal=jpcc,manuscript=article]{achemso}
\usepackage{achemso}
\setkeys{acs}{usetitle = true}
\usepackage[version=3]{mhchem} % Formula subscripts using \ce{}
\usepackage[T1]{fontenc}       % Use modern font encodings
\usepackage{color,soul}
\usepackage{amsmath,amssymb}
\author{Shikha Saini, Pooja Basera, Manish Kumar, Preeti Bhumla, Saswata Bhattacharya} 
\affiliation{Department of Physics, Indian Institute of Technology Delhi, Hauz Khas, New Delhi, 110016, India}
\email{saswata@physics.iitd.ac.in [SB]}
\phone{+91-2659 1359}
\fax{+91-2658 2037}

\title[An \textsf{achemso} demo]
{Metastability Triggered Reactivity in Clusters at Realistic Conditions: A Case Study of N-doped (TiO$_2$)$_n$ for Photocatalysis}

\begin{document}
%\begin{tocentry}

%Some journals require a graphical entry for the Table of Contents.
%This should be laid out ``print ready'' so that the sizing of the

%Inside the \texttt{tocentry} environment, the font used is Helvetica
%8\,pt, as required by \emph{Journal of the American Chemical

%The surrounding frame is 9\,cm by 3.5\,cm, which is the maximum
%permitted for  \emph{Journal of the American Chemical Society}
%graphical table of content entries. The box will not resize if the
%content is too big: instead it will overflow the edge of the box.

%This box and the associated title will always be printed on a
%separate page at the end of the document.

%\end{tocentry}

\begin{abstract} \label{abstract}
Here we report a strategy, by taking a prototypical model system for photocatalysis (viz. N-doped (TiO$_2$)$_n$ clusters), to accurately determine low energy metastable structures that can play a major role with enhanced catalytic reactivity. Computational design of specific metastable photocatalyst with enhanced activity is never been easy due to plenty of isomers on potential energy surface. This requires fixing various parameters viz. (i) favorable formation energy, (ii) low fundamental gap, (iii) low excitation energy and (iv) high vertical electron affinity (VEA) and low vertical ionization potential (VIP). We validate here by integrating several first principles based methodologies that consideration of the global minimum structure alone can severely underestimate the activity. As a first step, we have used a suite of genetic algorithms [viz. searching clusters with conventional minimum total energy ((GA)$_\textrm{E}$); searching clusters with specific property i.e. high VEA ((GA)$_\textrm{P}^{\textrm{EA}}$), and low VIP ((GA)$_\textrm{P}^{\textrm{IP}}$)] to model the N-doped (TiO$_2$)$_n$ clusters. Following this, we have identified its free energy using ab initio thermodynamics to confirm that the metastable structures are not too far from the global minima. By analyzing a large dataset, we find that N-substitution ((N)$_\textrm{O}$) prefers to reside at highly coordinated oxygen site to maximize its coordination, whereas N-interstitial ((NO)$_\textrm{O}$) and split-interstitial ((N$_2)_\textrm{O}$) favor the dangling oxygen site. Interestingly, we notice that each types of defect (viz. substitution, interstitials) reduce the fundamental gap and excitation energy substantially. However, (NO)$_\textrm{O}$ and (N$_2)_\textrm{O}$ doped clusters are the potential candidates for overall water splitting, whereas N$_\textrm{O}$ is congenial only for oxygen evolution reaction. 
\end{abstract}
\section{Introduction}\label{Introduction}
Accurate prediction of the structure of clusters as a catalyst, under reaction conditions, is the most fundamental challenge to get a detailed understanding of the active sites and their importance. %Some obvious questions arise naturally, for example, ``which are the species present in the real catalyst and what are their structures?'' 
Determining the catalyst structures at various reaction conditions is still a great challenge even for modern experimental methods. First principles based state-of-the-art global optimization methods viz. genetic algorithm (GA)\cite{alexandrova2005search,kanters2014cluster, davis2015birmingham, erlebach2015structure, bandow2006larger, chen2017modeling}, basin-hopping (BH)\cite{wales1997global}, parallel tempering\cite{sambridge2013parallel}, particle-swarm optimization (PSO)\cite{call2007global}, stochastic tunneling\cite{wenzel1999stochastic}, simulated annealing (SA)\cite{wang2009structural} etc. can predict the catalysts' structure. Moreover, if we know some primary information such as the elemental constituents in the catalyst and its reaction conditions (e.g. temperatures and pressures, doping concentration, etc.), accurate prediction of the equilibrium state is in principle possible via \textit{ab initio} thermodynamics~\cite{bhattacharya2014efficient}. However, this situation becomes complicated for real catalysis if we look deep into practical correlation of the predicted structure and its relevance with the concerned catalytic reactivity. The structures closer to the global minimum (based on ground state total energies) have higher occurrence probability at a finite temperature, but that does not ensure these structures are responsible for the observed activity~\cite{sun2018metastable}. On the other hand, metastable isomers of the catalyst are definitely not as stable as its global minimum but may lead to having higher activity due to presence of active sites. Moreover, under reaction conditions, the catalyst comprises of a wide range of structures all of which could be active to some extent in the catalytic reaction~\cite{Alexandrova2017ACS}. Over the past, in most of the theoretical studies of clusters, it is assumed that experimentally probed clusters are in their ground state conditions due to thermodynamic equilibrium~\cite{reuter2005ab,reuter2001composition, reuter2004oxide, saini2018structure, basera2019stability}. Contrary to this, few experimental and theoretical studies have revealed that the experimentally detected clusters are the metastable isomers, rather than the ground-state ones~\cite{kronik2002highest, saini2019unraveling, NoaPRL2012,SaswataPRBR2015}. Here we present a robust theoretical approach to study the active sites of a cluster by taking a prototypical model system for Transition-metal (TM) oxides, in particular titanium dioxide (TiO$_2$), owing to its ubiquity, low cost, stability, nontoxicity, catalytic activity and environment friendly nature.

TiO$_2$ has a great significance in photocatalysis from the perspective of industrial applications~\cite{s1,shi2012synthesis, sulaiman2018effects, s2, kapilashrami2014probing, s3, pooja2019RSC, lamiel2017predicting, pan2013defective, s6}. However, the wide band gap of TiO$_2$ that only absorbs the UV light of the solar spectrum, limits its efficiency in technological applications. Previous works suggest that the non-metal doping enhances the photoactivity of nanoclusters of TiO$_2$~\cite{asahi2001visible, burda2003enhanced, gole2004highly, mowbray2009stability, chen2008electronic, chen2004photoelectron}. The higher photocatalytic activity and stronger optical response are noticed with an increase of the N-doping concentration in TiO$_2$ nanophotocatalyst~\cite{cong2007synthesis,r1, burda2003enhanced,r2, chen2008electronic, irie2003nitrogen,r3, r4, r5, r6, r7}. Xiaobo \textit{et al.} have revealed additional electronic states for non-metal dopants (N, C and S), above the valence band edge of pure TiO$_2$ nanomaterials using X-ray photoelectron spectroscopy (XPS)~\cite{chen2008electronic}, which lead to the substantial modification in the optical properties. On the contrary, in theoretical study of Shevlin \textit{et al.}, no response in visible region is reported for N-doped TiO$_2$ clusters\cite{shevlin2010electronic}. The viable reason of this discrepency of theoretical and experimental finding is, in the theoretical work they have addressed only the most stable substitutional N-defects, whereas in the XPS spectra both substitutional and interstitial (meta)stable defects are detected~\cite{chen2008electronic}. 

\begin{figure}[h!]
	\includegraphics[width=0.8\textwidth]{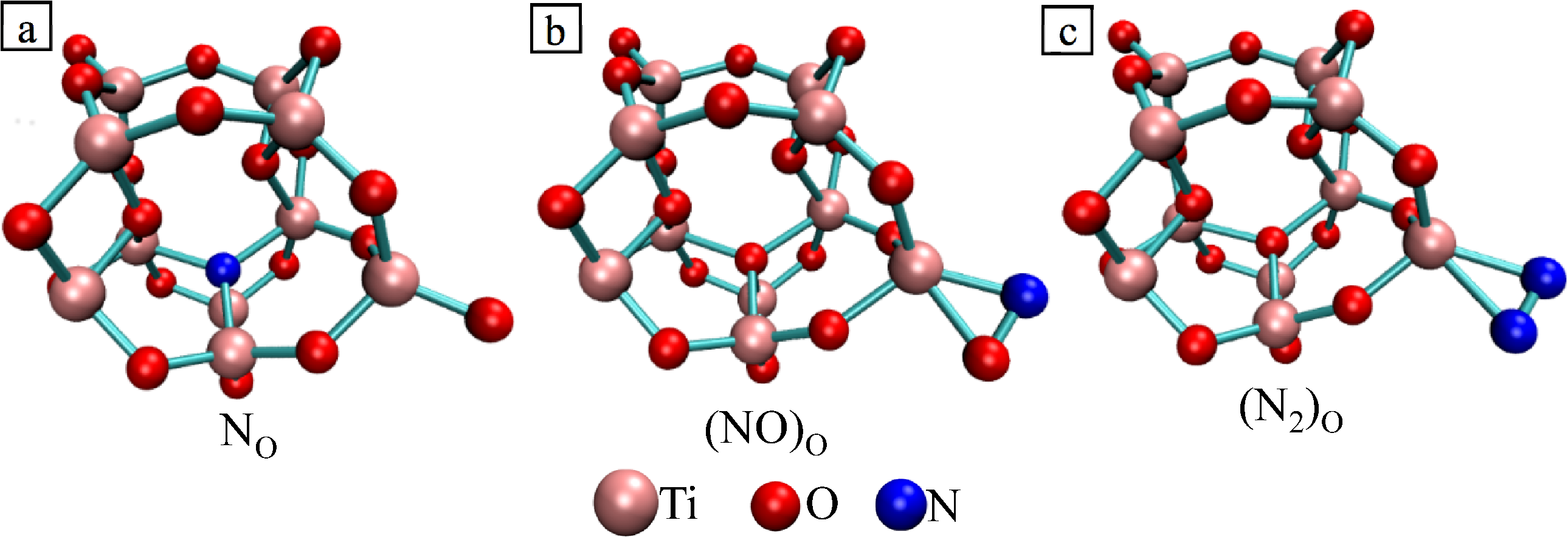}
	\caption{Energetically preferable position of different types of defect in N-doped (TiO$_2$)$_{10}$ cluster: (a) N$_\textrm{O}$, (b) (NO)$_\textrm{O}$ and (c) (N$_2)_\textrm{O}$.}
	\label{fig1}
\end{figure}

Therefore, it's well known these days that electronic properties of nanoclusters highly depend on structural configuration and particularly for the catalysis purpose, metastable structures are promising choice rather than the global minimum structure~\cite{kronik2002highest, sun2018metastable, NoaPRL2012, saini2018structure}. Note that previous studies have suggested that clusters possessing a high vertical electron affinity (VEA) or a low vertical ionization potential (VIP) are the promising choice as a photocatalyst. This is due to their ability to accept or donate an electron more readily~\cite{NoaPRL2012, SaswataPRBR2015, kronik2002highest}. We have, therefore, implemented a suite of massively parallel cascade genetic algorithms (GA). The first is the conventional energy-based GA [viz. (GA)$_\textrm{E}$] as described in detail in Ref~\cite{bhattacharya2014efficient}. This will give us all the local isomers close to energy based global minimum. The second GA is tailored explicitly to find metastable structures having specific bias for a property [viz. (GA)$_\textrm{P}$]. This specific property is used to evaluate the fitness function for (GA)$_\textrm{P}$. If this property is high vertical electron affinity, we call it (GA)$_\textrm{P}^{\textrm{EA}}$, whereas if it's low vertical ionization potential we represent it (GA)$_\textrm{P}^{\textrm{IP}}$. As a test case we have shown the performance of these three GAs [viz. (GA)$_\textrm{E}$, (GA)$_\textrm{P}^{\textrm{EA}}$, (GA)$_\textrm{P}^{\textrm{IP}}$ ] for pristine (TiO$_2$)$_n$ clusters at various sizes in Fig S1 of supporting information (SI). It's clearly shown that while (GA)$_\textrm{E}$ searches low energy clusters, (GA)$_\textrm{P}$ focusses some metastable part of the PES to optimize some specific properties.  More details and validity of this implementation can be found in Ref~\cite{SaswataPRBR2015}. Here, we have applied these three GAs to build a database of pristine as well as doped (TiO$_2$)$_n$ clusters with n = 4 -- 10, 15, 20. Note that we have investigated three different configurations of N-doped (TiO$_2$)$_n$ nanoclusters to modify  electronic properties at sub-nanometer scale: (a) N replaces O-atom making a substitutional defect (N$_\textrm{O}$), (b) N as interstitial (NO)$_\textrm{O}$, and (c) (N$_2)_\textrm{O}$ where both N substitution as well as interstitial share the same site (as shown in Fig~\ref{fig1}). 

In this article, as a first step from an exhaustive scanning, we have considered three types of (un)doped (TiO$_2$)$_n$ clusters: (i) clusters having the minimum ground state total energy (ii) clusters with high vertical electron affinity (VEA), and (iii) clusters possessing the low vertical ionization potential (VIP). %We believe this helps us to build an exhaustive database of (meta)stable clusters. 
Note that despite (meta)stable structures are promising candidates for catalysis, their free energy of formation should not be too far away from the free energy based global minimum. Therefore, we determine the thermodynamic stability of these structures by minimizing its Gibbs' free energy of formation as a function of charge state at realistic conditions (e.g. temperature ($T$), oxygen partial pressure ($p_{\textrm O_2}$), doping)~\cite{bhattacharya2014efficient, g1, g2, g3, g4}. This facilitates us to estimate the probability of occurrence of these (meta)stable structures. Following this, a few clusters,  that are thermodynamically stable as well as possess active sites, are selected and their electronic structures are accurately analyzed using GW calculations. This is how we have systematically studied doped (TiO$_2$)$_n$ clusters for application in photocatalysis.
%%%%%%%%%%%%%%%%%%%%%%%%%%%%%%%

\section{Methodology}\label{Methodology}
All density functional theory (DFT) calculations have been performed using FHI-aims code, which is an all electron code with numeric, atom-centered basis set~\cite{blum2009ab}. To find the preferred site for different types of defects [N$_\textrm{O}$, (NO)$_\textrm{O}$ and (N$_2)_\textrm{O}$] in the clusters, we have employed cascade GA~\cite{bhattacharya2014efficient,SaswataPRBR2015}. Within our cascade GA approach~\cite{bhattacharya2014efficient,SaswataPRBR2015} successive steps employ increasingly more accurate level of theories and each of the next level takes information obtained from its immediate lower level. This way, structural information is passed between steps of the cascade, and certain unfit structures are filtered out. We have thoroughly benchmarked and tested the efficiency of our cascade GA to accelerate the evolution of structures. More details can be found in Ref\cite{bhattacharya2014efficient}. 
While running GA, the optimization is done with vdW-corrected~\cite{tkatchenko2009accurate} PBE~\cite{perdew1996generalized} functional [PBE+vdW]. We have used ``tight - tier 2'' settings~\cite{blum2009ab}, and force tolerance is set to 10$^{-5}$ eV/${\textrm \AA}$. %The van-der-Waals correction is calculated as implemented in Tkatchenko-Scheffler scheme~\cite{tkatchenko2009accurate}. 
We have  reported in our previous studies~\cite{bhattacharya2013stability, saini2018structure, bhattacharya2014efficient, saini2019unraveling, basera2019stability} that PBE+vdW energetics give qualitatively wrong prediction for stability of oxide systems. Therefore, inside cascade GA, right after optimization, we run a single point energy calculation via vdW-corrected-PBE0~\cite{perdew1996rationale} hybrid exchange correlation functional (PBE0+vdW), with ``tight - tier 2'' settings to determine the fitness function of various defect configurations. Note that for property based cascade GA i.e. (GA)$_\textrm{P}$ the VEA/VIP values are obtained via delta-scf method~\cite{SaswataPRBR2015} at the level of PBE0+vdW. Note that accurate choice of functional for estimation of the fitness function of different clusters is essential for a meaningful scanning of the PES. We have thoroughly validated this in Ref~\cite{bhattacharya2013stability, bhattacharya2014efficient, SaswataPRBR2015}.

After getting the low energy N-doped (TiO$_2$)$_n$ (n = 4 -- 10, 15, 20) configurations for all the sets of clusters ((GA)$_\textrm{E}$, (GA)$_\textrm{P}$$^\textrm{EA}$ and (GA)$_\textrm{P}$$^\textrm{IP}$ based clusters) (one case is shown in Fig S2 of SI), we study the thermodynamic stability of the N-doped (TiO$_2$)$_n$ clusters in an oxygen atmosphere using the \textit{ab initio} atomistic thermodynamics (\textit{ai}AT) approach. The concept of \textit{ai}AT has been initially developed for bulk semiconductors\cite{scheffler1986resonant,scheffler1988parameter}, surface oxide\cite{wang1998hematite,lee2000gaas,reuter2003composition,reuter2005ab} and later successfully applied to the gas phase clusters in the reactive atmosphere\cite{bhattacharya2013stability,bhattacharya2014efficient}. Using this approach, we can determine the stability of defect states 
by minimizing the Gibbs' free energy of formation at different $T$ and $p_{\textrm O_2}$. 
We have used many-body perturbation theory within the GW approximation to evaluate the fundamental gap (E$_g$) and excitation energy (E$_x$) of all the defect configurations. ``Really-tight'' numerical settings and tier 4 basis set~\cite{blum2009ab} are used to calculate E$_g$ at the level of G$_0$W$_0$@PBE0.

\section{Results and Discussions}\label{Results and Discussions}
\subsection{Structural details to form (N)$_\textrm{O}$, (NO)$_\textrm{O}$ and (N$_2$)$_\textrm{O}$ in (TiO$_2$)$_n$ clusters}
As mentioned above, to study N-doped (TiO$_2$)$_n$ clusters, as a first step, a robust scanning of the potential energy surface (PES) is done using (GA)$_\textrm{E}$, (GA)$_\textrm{P}^{\textrm{EA}}$, and (GA)$_\textrm{P}^{\textrm{IP}}$. This helps in building a large data set (global minimum and metastable clusters) to provide realistic description of structures and electronic properties. Moreover, it helps us to identify all possible energetically favorable positions of N related defects (either substitution, interstitial or combination of both) in (TiO$_2$)$_n$ clusters. In Fig~\ref{fig1}(a-c), we have shown the favorable defect sites as (N)$_\textrm{O}$, (NO)$_\textrm{O}$ and (N$_2$)$_\textrm{O}$ in N-doped (TiO$_2$)$_{10}$ cluster, respectively. Relative energies of different N-doped configurations [viz. (N)$_\textrm{O}$, (NO)$_\textrm{O}$, (N$_2$)$_\textrm{O}$] at different O-sites of (TiO$_2$)$_{10}$ cluster are shown in SI (see the Fig S2) as a test case.

We have found that O-site with high coordination is preferable for substitutional defect. This finding is in good agreement with the previous simulation study~\cite{shevlin2010electronic}. Note that from the perspective of electronic configuration, N-atom has three unpaired electrons in the outermost shell, whereas O-atom has only two electrons. Therefore, the N-atom will favor more folded O-site to attain the maximum coordination number in the clusters. For interstitial case, we have observed that interstitial N-atom prefers to form the bond with dangling O-atoms because these oxygen atoms have the localized states at the HOMO level (see Fig S3). The interaction of (NO)$_\textrm{O}$ doping affects the HOMO states, which may lead to substantial modification in their electronic properties. Moreover, since N-atom has unpaired electrons, it has tendency to share the electrons with more electronegative atom (dangling O-atom is having less coordination number). Likewise (NO)$_\textrm{O}$, (N$_2$)$_\textrm{O}$ also favors the dangling O-site (see Fig~\ref{fig1}). By getting various cluster configurations from GA, we determine the thermodynamic stability of the clusters at finite  $T$, $p_{\textrm O_2}$ as discussed in the following section.

\subsection{Thermodynamic stability of (N)$_\textrm{O}$, (NO)$_\textrm{O}$ and (N$_2$)$_\textrm{O}$ in (TiO$_2$)$_n$ clusters}
We address the thermodynamic stability of different types of charged defects ([(N)$_\textrm{O}$]$^q$, [(NO)$_\textrm{O}$]$^q$ and [(N$_2$)$_\textrm{O}$]$^q$ with $q$ = -2, -1, 0, +1, +2) in an oxygen atmosphere using \textit{ai}AT. The phase diagrams are obtained by determining the Gibbs' free energy of formation of all N-doped clusters as a function of $T$, $p_{\textrm O_2}$ and chemical potential of electron ($\mu_\textrm{e}$) by using the following equations: \\
(i) The formation energy of (N)$_\textrm{O}$ in the charge state $q$ is given by: 
\begin{equation}
\begin{split}
\textrm{E}_\textrm{f}(\textrm{N}_\textrm{O}^{q}) = \textrm{E[NTi}_{n}\textrm{O}_{2n-1}]^{q} - \textrm{E[Ti}_{n}\textrm{O}_{2n}]^0 + \mu_\textrm{O} - \mu_\textrm{N} + q\mu_\textrm{e}
\end{split}
\end{equation}
(ii) The formation energy for interstitial N i.e. (NO)$_\textrm{O}$ is:
\begin{equation}
\begin{split}
\textrm{E}_\textrm{f}((\textrm{NO})_\textrm{O}^{q}) = \textrm{E[NTi}_{n}\textrm{O}_{2n}]^{q} - \textrm{E[Ti}_{n}\textrm{O}_{2n}]^0 - \mu_\textrm{N}+q\mu_\textrm{e}
\end{split}
\end{equation}
(iii) Similarly, the formation energy for (N$_2$)$_\textrm{O}$ can be written as:
\begin{equation}
\begin{split}
\textrm{E}_\textrm{f}((\textrm{N}_2)_\textrm{O}^q) = \textrm{E[N}_2\textrm{Ti}_{n}\textrm{O}_{2n-1}]^{q} - \textrm{E[Ti}_{n}\textrm{O}_{2n}]^0 + \mu_\textrm{O}
- 2\mu_\textrm{N}+ q\mu_\textrm{e}\\  
\end{split}
\end{equation}
where, $\mu_\textrm{N} = \Delta\mu_\textrm{N}+\frac{1}{2}\textrm{E[N}_2]+\frac{h\nu_{NN}}{4}$ and
$\mu_\textrm{O} = \Delta\mu_\textrm{O} + \frac{1}{2}\textrm{E[O}_2]+\frac{h\nu_{OO}}{4}$. 
Here, $\textrm{E[NTi}_{n}\textrm{O}_{2n-1}]^{q}$, $\textrm{E[NTi}_{n}\textrm{O}_{2n}]^{q}$, $\textrm{E[N}_2\textrm{Ti}_{n}\textrm{O}_{2n-1}]^{q}$ and $\textrm{E[Ti}_{n}\textrm{O}_{2n}]^0$ are the total energies corresponding to (N)$_\textrm{O}$, (NO)$_\textrm{O}$, (N$_2$)$_\textrm{O}$ doped and undoped clusters, respectively. $\textrm{E[O}_2]$ and $\textrm{E[N}_2]$ are the total energies of O$_2$ and N$_2$ molecules, respectively. $\mu_\textrm{e}$ is varied from valence band maximum to conduction band minimum of the bulk TiO$_2$. $\nu_{OO}$ and $\nu_{NN}$ are the stretching frequencies of O--O and N--N bonds, respectively. The formation energies of charged and neutral defects depend on $\mu_\textrm{O}$, which incorporates the effect of $T$ and $p_{\textrm O_2}$. $\Delta\mu_\textrm{O}$ as a function of $T$ and $p_{\textrm{O}_2}$ is calculated as follow~\cite{bhattacharya2014efficient}:
\begin{equation}\begin{split}
\Delta\mu_\textrm{O}(T, p_{\textrm{O}_2}) &= \frac{1}{2}\left[ -k_\textrm{B}T \ln\left[\left(\frac{2\pi m}{h^2}\right)^\frac{3}{2}\left(k_\textrm{B}T\right)^\frac{5}{2}\right]\right.
\\
&\quad+ k_\textrm{B}T \ln p_{\textrm{O}_2} - k_\textrm{B}T \ln \left(\frac{8\pi^2I_Ak_\textrm{B}T}{h^2}\right)
\\
&\quad+ k_\textrm{B}T \ln \left[1-\exp\left(\frac{-h\nu_\textrm{OO}}{k_\textrm{B}T}\right)\right]
\\
&\quad\left.-  k_\textrm{B}T \ln \mathcal{M} + k_\textrm{B}T \ln \sigma \right]
\end{split}\end{equation}
where $m$ is the mass, $I_A$ is the moment of inertia of $\textrm{O}_2$ molecule, $\mathcal{M}$ is the spin multiplicity and $\sigma$ is the symmetry number. Similarly, $\mu_\textrm{N}$ is estimated by the formation of N$_2$ molecule, i.e., $\Delta\mu_\textrm{N}$ = -0.25 eV at ambient condition~\cite{basera2019stability, kumar2019photocatalytic}.

\begin{figure}[h!]
	\includegraphics[width=0.9\textwidth]{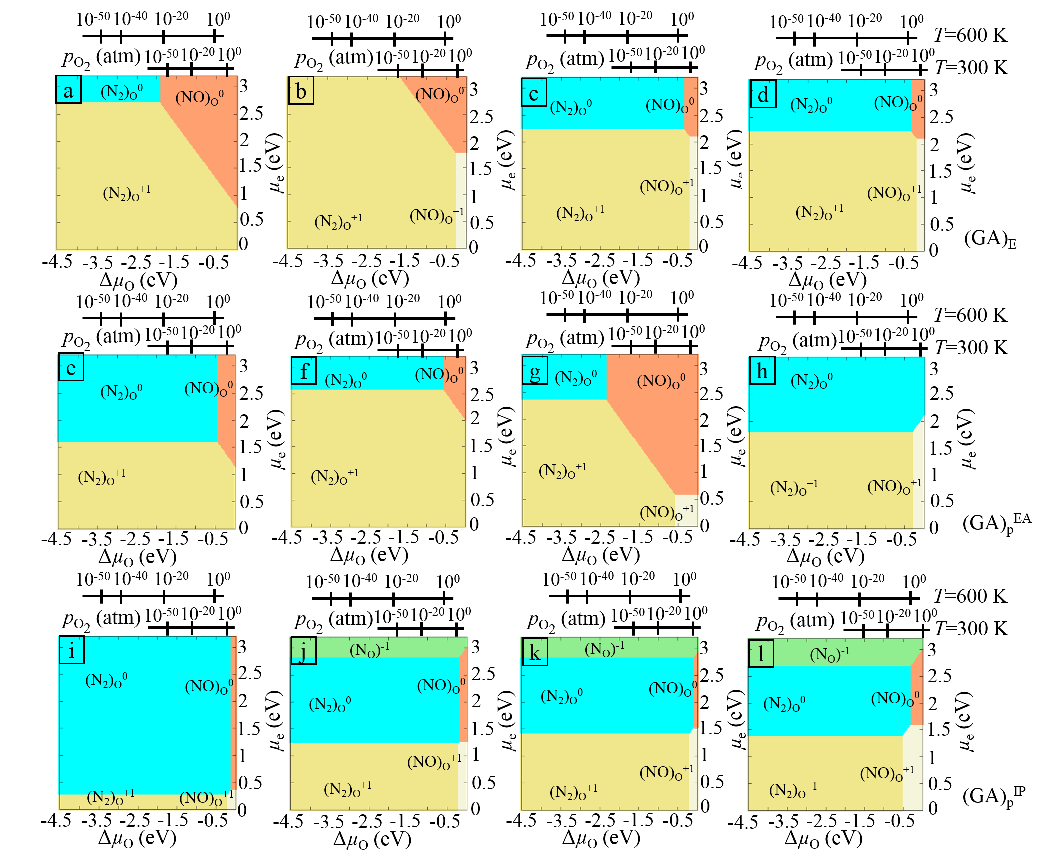}
	\caption{(color online) 2D view of 3D phase diagrams obtained for N-doped (GA)$_\textrm{E}$ [upper panel], (GA)$_\textrm{P}$$^\textrm{EA}$ [middle panel] and (GA)$_\textrm{P}$$^\textrm{IP}$ [lower panel] clusters in different charge states for size n = 5 (a, e, i), 10 (b, f, j), 15 (c, g, k) and 20 (d, h, l). Colored regions show the most stable defect states at realistic conditions ($T$, $p_{\textrm O_2}$ and $\mu_\textrm{e}$). The top axes are representing the pressure scale of O$_2$ at $T$ = 300K and 600K.}
	\label{fig3}
\end{figure}

3D phase diagrams of (GA)$_\textrm{E}$ clusters for size n = 5, 10, 15, 20 are shown in Fig~\ref{fig3}(a), ~\ref{fig3}(b),~\ref{fig3}(c) and~\ref{fig3}(d), respectively. At lower values of $\mu_\textrm{e}$ (p-type doping),  (N$_2$)$_\textrm{O}^{+1}$ defect is predominant for a wide range of $\Delta\mu_\textrm{O}$. However, (NO)$_\textrm{O}^{+1}$ defect is also observed at higher range of pressure for size n = 10, 15, 20. At higher values of $\mu_\textrm{e}$ (n-type doping), (N$_2$)$_\textrm{O}^{0}$ defect is stable at lower range of pressure except for n = 10 case. However at feasible pressure, (NO)$_\textrm{O}^{0}$ defect is found to be the most stable. Hence, we can summarize that interstitial defects [(NO)$_\textrm{O}$ and (N$_2$)$_\textrm{O}$] are most stable in (GA)$_\textrm{E}$ based (TiO$_2$)$_n$ clusters. However, substitutional defect N$_\textrm{O}$ has higher formation energy
at ambient condition ($T$ = 300 K, $p_{\textrm O_2}$ = 1 atm), that can be seen clearly in 2D phase diagrams as shown in SI (Fig S4(a-d)).
Note that until date, theoretical calculations are limited to non-metal doping (that to substitution only) in TiO$_2$ clusters closer to energy based global minimum~\cite{mowbray2009stability, shevlin2010electronic}. It is therefore of profound interest to address the stability and electronic structure of property based doped clusters (i.e. clusters generated from (GA)$_\textrm{P}$$^\textrm{EA}$, and (GA)$_\textrm{P}$$^\textrm{IP}$ algorithms). The latter may have a better correlation with the experimentally detected clusters.
%In the previous simulation studies of non-metal doped TiO$_2$ clusters only substitutional defect is considered.  
Likewise (GA)$_\textrm{E}$ clusters, the property based clusters follow almost the same trend at lower values of $\mu_\textrm{e}$ [see Fig~\ref{fig3}e-l]. However, (N$_2$)$_\textrm{O}^{0}$ phase is most probable in n-type region (higher $\mu_\textrm{e}$), whereas a slight portion of (NO)$_\textrm{O}^0$ defect is apparent at high pressures for n = 5, 10 and 15 cases (see Fig ~\ref{fig3}(e-h)).  
Interestingly, we notice that substitutional phase i.e., N$_\textrm{O}^{-1}$ is only visible in n-type doping region along with interstitial phases for doped clusters (scanned via (GA)$_\textrm{P}$$^\textrm{IP}$) having size n = 10, 15, 20 (see Fig ~\ref{fig3}(j-l)). 
We can make a key conclusion from Fig~\ref{fig3} that N interstitial defects ((N$_2$)$_\textrm{O}^{0/+1}$, (NO)$_\textrm{O}^{0/+1}$) are most prominent at a given $T$ and $p_{\textrm O_2}$ in N-doped TiO$_2$ clusters. Further, we have explored fundamental gap and excitation energy of (un)doped TiO$_2$ clusters to see their applicability in photocatalysis.

\subsection{Fundamental gap and excitation energy of (un)doped (TiO$_2$)$_n$ clusters}
Next, setting formation energy of global minimum at 0 eV of the respective class of configurations (viz. pristine, N$_\textrm{O}$, (NO)$_\textrm{O}$ and (N$_2$)$_\textrm{O}$), we have considered all the structures within an energy window of 3~eV generated via (GA)$_\textrm{E}$, (GA)$_\textrm{P}$$^\textrm{EA}$ and (GA)$_\textrm{P}$$^\textrm{IP}$ for further analysis of their electronic structure. The number 3~eV is chosen assuming this is large enough window for consideration of metastable isomers and anything beyond this is very less likely to appear in real experiments. 

\begin{figure}[h!]
	\includegraphics[width=0.8\textwidth]{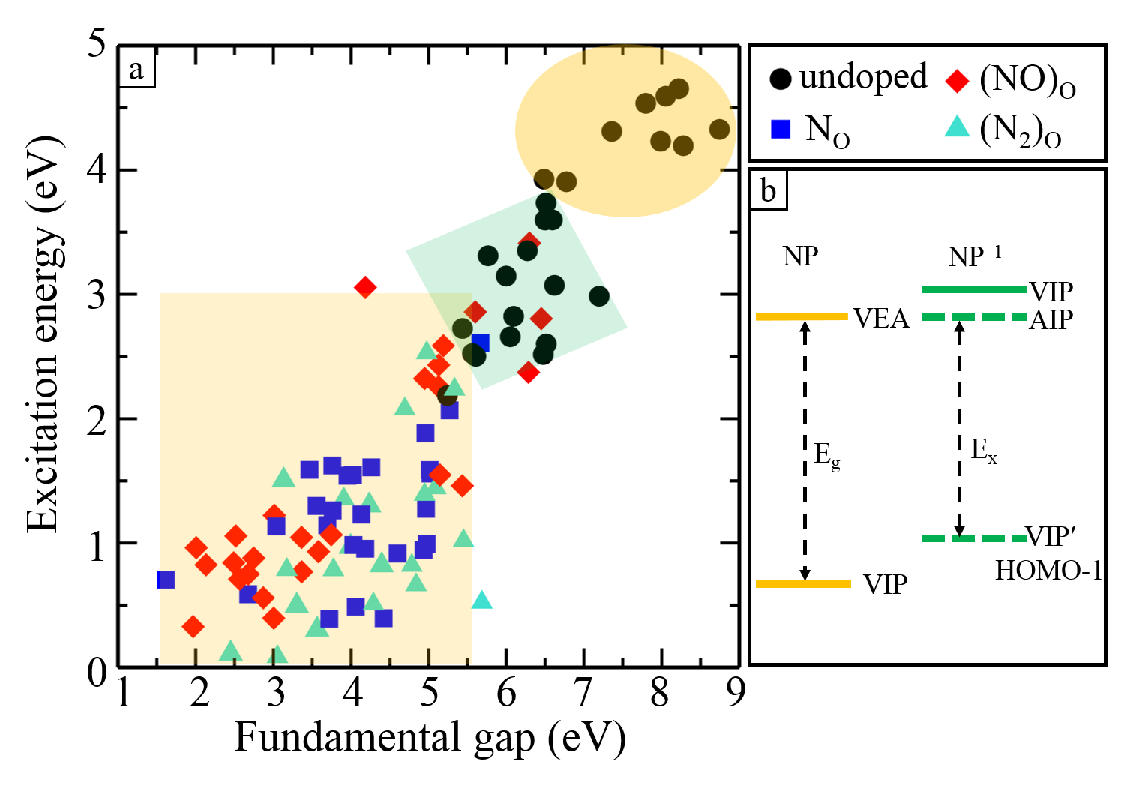}
	\caption{(color online) (a) fundamental gap (E$_g$) vs. excitation energy (E$_x$) (G$_0$W$_0$@PBE0) for all the (un)doped (TiO$_2$)$_n$ [n = 4 - 10, 15, 20] clusters. (b) In the schematic, yellow lines refer to the vertical ionization potential and vertical electron affinity of the neutral cluster (NP), while green line refers to the vertical ionization potential of negatively charged cluster. The dashed green lines define VIP$^{'}$ (the energetic cost of extracting an electron from the HOMO-1 level of the NP$^{-1}$) and AIP (adiabatic ionization potential corresponds to HOMO level of the NP$^{-1}$).}
	\label{fig4}
\end{figure}

We have then determined the fundamental gap (E$_g$) and excitation energy of all the N-doped (TiO$_2$)$_{n}$ clusters and compared with pristine counterpart (see Fig~\ref{fig4}a). These are computed at the level of G$_0$W$_0$@PBE0. Note that the difference of vertical electron affinity (VEA) and vertical ionization potential (VIP) gives the fundamental gap\cite{r49}. We can also define it as the energy required to make a pair of free charge carriers i.e. quasiparticle gap. If a particular cluster has simultaneously high vertical electron affinity (VEA) and low vertical ionization potential (VIP), it possesses the low fundamental gap (VIP -- VEA). Furthermore, this candidate would be the very active cluster as it can accept or donate an electron readily ~\cite{SaswataPRBR2015,marom2012structure, saini2018structure, morales2019understanding}. 
It is experimentally reported that photoelectron spectroscopy  of negatively charged clusters gives the information about the energy gap between the highest occupied molecular orbital (HOMO) and lowest unoccupied molecular orbital (LUMO) for neutral clusters \cite{zhai2007probing}. Therefore, the excitation energy for neutral species can be determined by their negatively charged species. The extra electron in the anion cluster occupies the LUMO of the neutral cluster, and thus, yields the first band peak, whereas the next peak of the band corresponds to ionization energy of the HOMO-1 of charged cluster. 
Therefore, to calculate the excitation energy of (un)doped clusters, we have taken the difference of ionization potential of HOMO-1 level of anion cluster and electron affinity of the neutral clusters as shown in schematic diagram in Fig~\ref{fig4}b.
Note that, the adiabatic ionization potential (AIP) of the anion cluster defines the vertical electron affinity (VEA) of the neutral clusters. We notice that experimental results of undoped clusters are consistent with the clusters generated via (GA)$_\textrm{P}$ rather than (GA)$_\textrm{E}$ (see Table S1). This means these are metastable isomers and any conventional total energy based global minumum search algorithm will miss them to detect. 

In Fig~\ref{fig4}a, we have shown the fundamental gap vs. excitation energy of (un)doped clusters for each set ((GA)$_\textrm{E}$, (GA)$_\textrm{P}$$^\textrm{EA}$ and (GA)$_\textrm{P}$$^\textrm{IP}$). The region enclosed by circle corresponds to the undoped (GA)$_\textrm{E}$ clusters, having the large values for fundamental gap and excitation energy. Note that, the same undoped clusters scanned via (GA)$_\textrm{P}$ show the lower values as compared to the (GA)$_\textrm{E}$ clusters (see enclosed  diamond shape region). As the undoped clusters possess the high excitation energy, this limits their applications for the photocatalysis. Interestingly, in N-doped clusters, each types of defect (viz. substitution, interstitials) reduce the fundamental gap (E$_g$) and excitation energy (E$_x$) substantially (see the enclosed area by box in Fig~\ref{fig4}a). The reduction in the fundamental gap and excitation energy could be ascribed to the presence of the dopant states, which lead to the new HOMO--LUMO levels in the clusters.

%%%
\begin{figure}[h!]
	\includegraphics[width=0.8\textwidth]{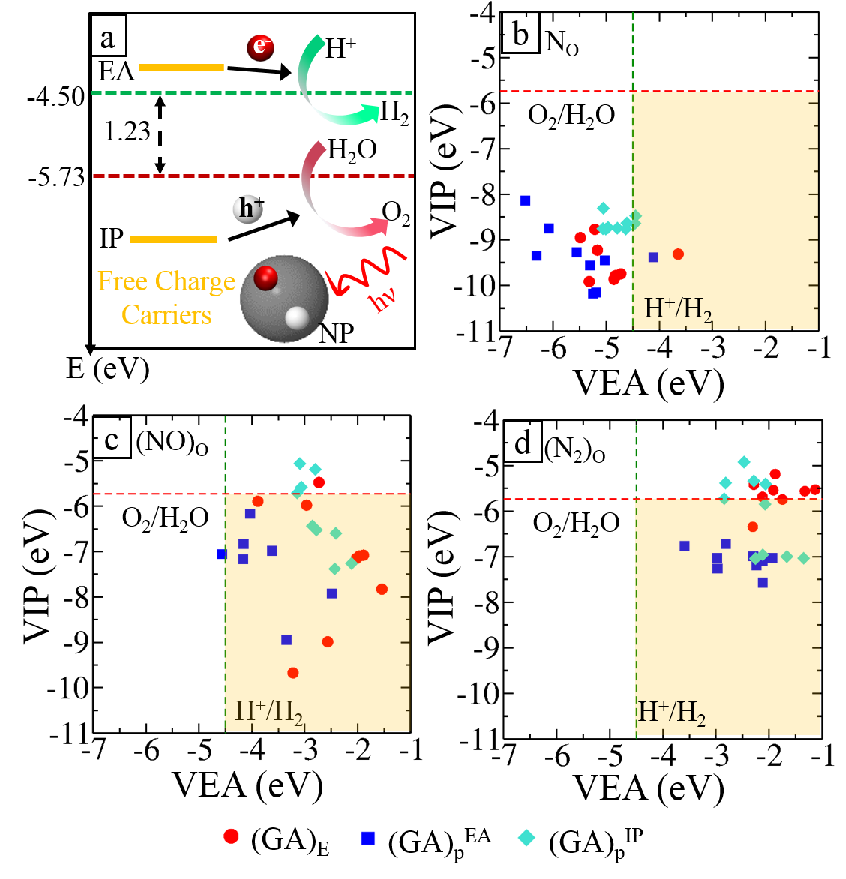}
	\caption{(a) The schematic diagram shows that how the (standard) reduction potentials (VIP and VEA) of the ideal photocatalyst (cluster) straddle the HER and OER potentials. The schematic shows the free charge carriers scenario, where the excited electron and hole are spatially separated within the particle due to negligible coulombic interaction. h$\nu$ defines the energy of the photon absorbed by the cluster. e$^-$ and h$^+$ stand for electron and hole, respectively. VIP refers to the cluster's ground-state ionization potential, whereas VEA to the ground-state electron affinity. VIP and VEA for doped clusters: (b) N$_\textrm{O}$, (c) (NO)$_\textrm{O}$, and (d) (N$_2$)$_\textrm{O}$. The dashed green and red lines represent the standard redox potentials for water reduction (H$^+$/H$_2$) and oxidation potential (O$_2$/H$_2$O) at pH = 0, respectively.} 
	\label{fig5}
\end{figure} 
\subsection{VIP and VEA of (un)doped clusters with respect to the water redox potentials}
Note that only reduction in fundamental gap can not assure the hydrogen generation via photocatalytic water splitting. The potential of free charge carriers (VIP, VEA) should have appropriate position to straddle the redox potentials of water for hydrogen evolution reaction (HER) and oxygen evolution reaction (OER).  
In order to analyze the ability of the clusters to drive the reduction of protons and the oxidation of water, four redox half-reactions are involved. Among these, two reactions are governed by exciton, and other two by free electron and hole\cite{berardo2015modeling,guiglion2014polymeric,butchosa2014carbon}.
In our case study, we have considered the latter one. Free charge carriers with the necessary chemical potential can in principle drive the water splitting half-reactions. The redox half-reactions are given below with the convention of reduction reactions\cite{guiglion2014polymeric}:
\begin{equation}
\textrm{NP} + \textrm{e}^{-} \leftrightarrows \textrm{NP}^{-1}
\end{equation}
\begin{equation}
\textrm{NP}^{+1} + \textrm{e}^{-} \leftrightarrows \textrm{NP}
\end{equation}
here NP is the neutral cluster, and $\textrm{NP}^{-1}, \textrm{NP}^{+1}$ represent the photocatalyst with a free electron in the conduction band and hole in valence band, respectively.
In equation 5 and 6, free electron act as a reductant and free hole will acts as an oxidant, respectively. 
The free energies of half-reactions are given in equation 7 and 8:
\begin{equation}
\Delta E\textrm{(6)} = E(\textrm{NP}^{-1}) - E(\textrm{NP}) = -\textrm{EA}
\end{equation}
\begin{equation}
\Delta E\textrm{(7)} = E(\textrm{NP}) - E(\textrm{NP}^{+1}) = -\textrm{IP}
\end{equation}
$\Delta$E(6) and $\Delta$E(7) are equal to negative of adiabatic electron affinity and ionization potential, respectively. 
Note that we have used vertical approximation that ignores the nuclear relaxation and yields vertical potential.
For water splitting photocatalyst, the VIP (energy required to extract an electron from the HOMO) level must be located below the OER potential (O$_2$/H$_2$O), whereas the VEA (the energy released while adding the electron to LUMO) level must be above the HER \textit{}potential (H$^+$/H$_2$) as shown in the schematic diagram of Fig~\ref{fig5}a.
Using the information of VIP and VEA, one can calculate the reduction potentials associated with the free charge carriers.  
Specifically, we use many body perturbation theory to calculate the thermodynamic driving force for the water splitting half-reactions 5 and 6. Experimental potential values are given relative to the Standard Hydrogen Electrode (SHE) (pH = 0). In practice, the required overall potential difference is larger than 1.23 eV to overcome energetic losses and kinetic barriers. In Fig~\ref{fig5}(b, c, d), we show the G$_0$W$_0$@PBE0 predicted vertical potentials relative to the SHE to obtain the potential candidates for
photocatalytic water splitting. The colored area (as in Fig~\ref{fig5}(b, c, d)) represents the suitable region for overall water splitting. For N$_\textrm{O}$ dopant case, all the candidates (generated via (GA)$_\textrm{E}$ and (GA)$_\textrm{P}$) have only the suitable VIP for OER as shown in Fig~\ref{fig5}(b). Hence, the N$_\textrm{O}$ doped clusters are not potential candidates for overall water splitting except two [(GA)$_\textrm{E}$ and (GA)$_\textrm{P}$$^\textrm{EA}$ of size 9 and 10, respectively]. In case of (NO)$_\textrm{O}$ doping, maximum points lie within suitable region for both the potentials (see Fig~\ref{fig5}(c)). In Fig~\ref{fig5}(d), for (N$_2$)$_\textrm{O}$ doping, all the (GA)$_\textrm{P}$$^\textrm{EA}$ clusters are inside the colored region whereas majority of (GA)$_\textrm{E}$ and few (GA)$_\textrm{P}$$^\textrm{IP}$ clusters are lying above the water oxidation potential level (O$_2$/H$_2$O). The latter are appropriate for HER.
Among different doped configurations, (NO)$_\textrm{O}$ is the promising choice for photocatalytic water splitting (see Fig~\ref{fig5}). Note that maximum isomers scanned via (GA)$_\textrm{P}$$^\textrm{EA}$ is found to be best for photocatalytic water splitting followed by isomers generated by (GA)$_\textrm{P}$$^\textrm{IP}$ and (GA)$_\textrm{E}$ for all the doped configurations. This further validates the importance of a dedicated algorithm viz. (GA)$_\textrm{P}$ to capture the metastability triggered reactivity. 

\subsection{Electronic structure of doped (TiO$_2$)$_{10}$ clusters}
\begin{figure}[h!]
	\includegraphics[width=0.9\textwidth]{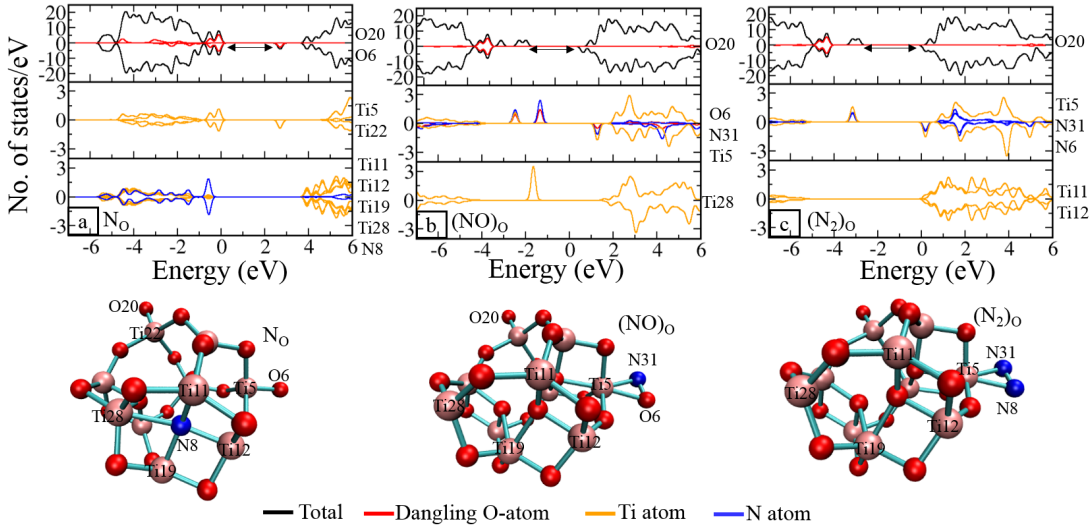}
	\caption{ TDOS and PDOS of (GA)$_\textrm{P}$$^\textrm{EA}$ based doped (TiO$_2$)$_{10}$ clusters: (a) N$_\textrm{O}$, (b) (NO)$_\textrm{O}$, (c) (N$_2)_\textrm{O}$. The respective electronic configuration of doped state is shown below the DOS. Double headed arrows are representing the HOMO--LUMO gap.}
	\label{fig6}
\end{figure}

To have in-depth understanding on the role of dopants in reducing the fundamental gap and their applicability in photocatalytic water splitting, we have analyzed the electronic density of states (DOS). 
The total DOS (TDOS) and partial DOS (PDOS) of (GA)$_\textrm{P}$$^\textrm{EA}$ based doped  (TiO$_2$)$_{10}$ clusters are shown in Fig~\ref{fig6}.  
Note that we have shown the PDOS of only those atoms, which have the major contribution at HOMO--LUMO levels.  In undoped (TiO$_2$)$_{10}$ cluster, orbitals of dangling O-atoms contribute to HOMO level, whereas the LUMO level is mainly attributed by the Ti-atoms that have the maximum distance from dangling O (see the Fig S3).
In case of N-subsitutional doping (N)$_\textrm{O}$, N is 4-folded and tightly bonded with Ti-atoms, that results in deep states far away from the Fermi-level  (see lower panel of Fig~\ref{fig6}a, S5a and S5d). Further, the charge density is calculated on N-site ($-$0.38) which is comparable to charge density on substitutional O-atom site ($-$0.40) of undoped case. This signifies that it will act as a deep donor site. We have also noticed the unoccupied deep mid gap states, which are associated with the dangling O-atoms and their bonded Ti-atoms. Consequently, the fundamental gap (E$_g$) is reduced in the N$_\textrm{O}$ doped structures, where N is highly coordinated to Ti (see in Fig~\ref{fig6}a). The aforementioned states shift the LUMO towards Fermi level and thus, N$_\textrm{O}$ doped clusters are not suitable for HER (see Fig S5a and S5d), which can also be observed from Fig~\ref{fig5}b.

In case of (NO)$_\textrm{O}$ doping, we find the occupied N-states above the HOMO level, which are overlapped with states of bonded atoms to the dopant as shown in Fig~\ref{fig6}b, S5b and S5e. Since oxygen is more electronegative than nitrogen, nitrogen might transfer charge to the bonded oxygen. As a result, dopant acts as a donor, and introduces occupied states near the HOMO. Therefore, in Fig S4b, for p-type doping region, the formation energy of (NO)$_\textrm{O}$ defect with +1 charge state is minimum. 
In few cases of (GA)$_\textrm{P}$$^\textrm{IP}$ based clusters, the manifestation of deep occupied mid gap states of Ti-atoms, deteriorates their oxidation potential (see the Fig S5e).
Similarly, for (N$_2)_\textrm{O}$ doping, N-atoms yield the occupied states, which are overlapped with states of bonded Ti as shown in Fig~\ref{fig6}c, S5c and S5f. (N$_2)_\textrm{O}$ transfers the charge to the Ti, which results in the strong bonding between N$_2$ entity and Ti-atom.  However, in (GA)$_\textrm{E}$ clusters, we observe the occupied mid gap states of Ti-atom in (N$_2)_\textrm{O}$ dopant case (see Fig S5c). As a result, their oxidation power is degraded. Therefore, all the points of (N$_2)_\textrm{O}$ dopant for (GA)$_\textrm{E}$ based clusters are lying above the OER potential in Fig~\ref{fig5}d. Hence, (N$_2)_\textrm{O}$ dopant in (GA)$_\textrm{E}$ clusters is not a desirable choice for overall water splitting. 
\begin{figure}[h!]
	\includegraphics[width=0.9\textwidth]{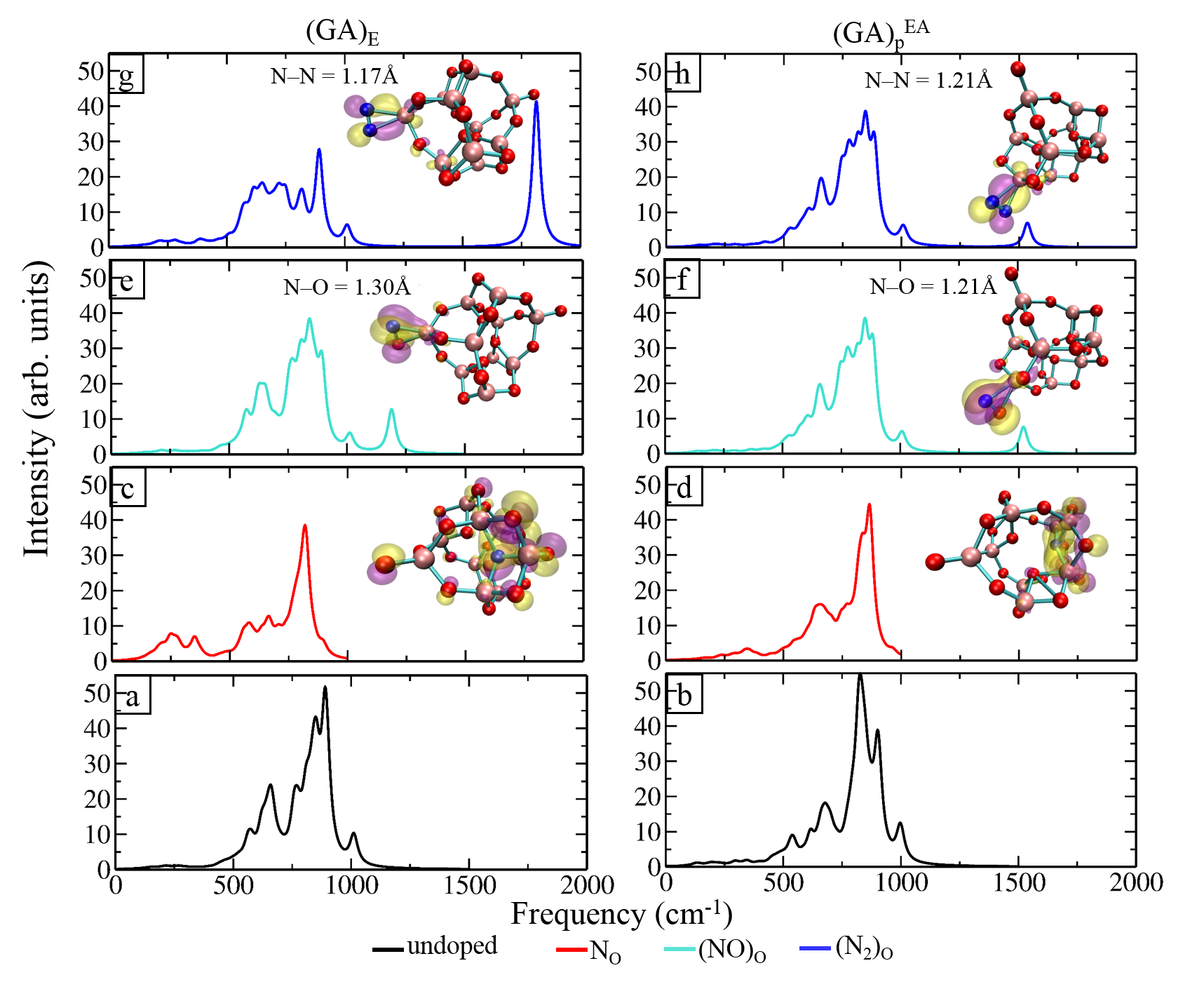}
	\caption{IR spectra of (GA)$_\textrm{E}$ (left column) and (GA)$_\textrm{P}$$^\textrm{EA}$ (right column) based (TiO$_2$)$_{10}$ clusters: (a, b) undoped, (c, d) N$_\textrm{O}$, (e, f) (NO)$_\textrm{O}$ and (g, h) (N$_2)_\textrm{O}$.}
	\label{fig7}
\end{figure}
In the next step, we have simulated the infrared (IR) spectra to determine the characteristic vibrational modes which are induced by the dopant sites in the clusters. For the concise description, we have considered repersentative cases as (GA)$_\textrm{E}$ and (GA)$_\textrm{P}$$^\textrm{EA}$ based clusters of size n = 10. In (N)$_\textrm{O}$ dopant cases, the spectrum has shifed to the lower frequency in comparison to the undoped cluster (see Fig~\ref{fig7}(a-d)). Contrary to this, in the IR spectra of (NO)$_\textrm{O}$ and (N$_2)_\textrm{O}$ doped clusters, we have noticed the additional peaks above the highest peak of undoped cluster as shown in Fig~\ref{fig7}(e-h). The N--O and N--N moieties are found to be responsible for the new emerging peaks in the spectra. In addition to this, we have seen the localized electron density at N--O and N--N moieties, which confirms that these moieties are the active centers in the clusters. However, in (N)$_\textrm{O}$ dopant cases the electron density is not localized at a particular site. These observations would be helpful for future experiments to recognize the different type of dopant sites in the (TiO$_2$)$_n$ clusters. 

\section{Conclusion}
In summary, we have presented a robust methodology to design clusters with desired properties: favorable formation energy, high vertical electron affinity (VEA), and low vertical ionization potential (VIP). For this purpose, we have implemented a suite of massively parallel cascade genetic algorithm to predict the accurate structures of N-doped (TiO$_2$)$_n$ (n = 4 -- 10, 15, 20) clusters viz. N$_\textrm{O}$, (NO)$_\textrm{O}$, (N$_2)_\textrm{O}$. From exhaustive scanning, we reveal that N-substitutional (N$_\textrm{O}$) dopant prefers to occupy oxygen site which is highly coordinated with Ti-atoms, whereas interstitial dopants [viz. (NO)$_\textrm{O}$ and (N$_2)_\textrm{O}$]  reside at the dangling oxygen site. Further, we have analyzed the thermodynamic stability of different doped configurations in various charge states at finite  $T$ and $p_{\textrm O_2}$ using \textit{ab initio} atomistic thermodynamics approach. We have found that (NO)$_\textrm{O}$ and (N$_2)_\textrm{O}$ are the most favorable phases in a wide range of $T$ and $p_{\textrm O_2}$. We have noticed significant reduction in the fundamental gap and excitation energy for doped clusters, which accounts for their application in photocatalysis. The favorable dopants for overall water splitting are (NO)$_\textrm{O}$ and (N$_2)_\textrm{O}$, whereas N$_\textrm{O}$ doped clusters are only suitable for OER. Moreover, (NO)$_\textrm{O}$ and (N$_2)_\textrm{O}$ possess the localized charge density at the dopant site. The relative efficiency for generating structures for overall water splitting is (GA)$_\textrm{P}$$^\textrm{EA}$ $>$ (GA)$_\textrm{P}$$^\textrm{IP}$ $>$ (GA)$_\textrm{E}$  for all the doped configurations. Therefore, to capture the metastability triggered reactivity, the adapted methodology will be helpful to design the rational nanoparticles for efficient photocatalysis. 

\section{Acknowledgement}\label{Acknowledgement}
SS acknowledges CSIR, India, for the senior research fellowship [grant no. 09/086(1231)2015-EMR-I]. PB acknowledges UGC, India, for the senior research fellowship [grant no. 20/12/2015(ii)EU-V]. MK acknowledges CSIR, India, for the senior research fellowship [grant no. 09/086(1292)/2017-EMR-I]. PB acknowledges UGC, India, for junior research fellowship [Ref. No.: 1392/(CSIR-UGC NET JUNE 2018)]. SB acknowledges the core research grant from SERB research grant, DST, India (grant no. CRG/2019/000647). We acknowledge the High Performance Computing (HPC) facility at IIT Delhi for computational resources. 
%\newpage
%\section{TOC graphic}
%\begin{figure}[h!]
%\includegraphics[width=0.50\columnwidth,clip]{TOC2.png}
%\end{figure}  

\bibliography{references1}% Produces the bibliography via BibTeX.
\end{document}

%\begin{tocentry}

%Some journals require a graphical entry for the Table of Contents.
%This should be laid out ``print ready'' so that the sizing of the

%Inside the \texttt{tocentry} environment, the font used is Helvetica
%8\,pt, as required by \emph{Journal of the American Chemical

%The surrounding frame is 9\,cm by 3.5\,cm, which is the maximum
%permitted for  \emph{Journal of the American Chemical Society}
%graphical table of content entries. The box will not resize if the
%content is too big: instead it will overflow the edge of the box.

%This box and the associated title will always be printed on a
%separate page at the end of the document.

%\end{tocentry}

\begin{figure}[h!]
	\includegraphics[width=1.0\columnwidth,clip]{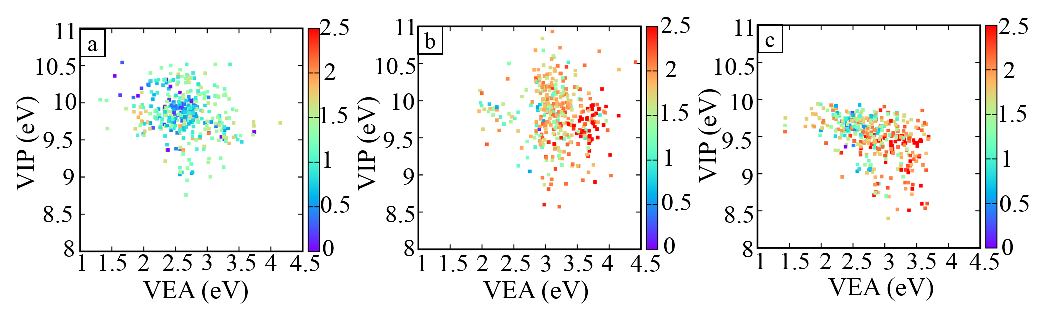}
	\caption{(color online) VIP vs VEA for the low-energy isomers of (a) (GA)$_\textrm{E}$, (b) (GA)$_\textrm{P}$$^\textrm{EA}$ and (c) (GA)$_\textrm{P}$$^\textrm{IP}$ based clusters. The color bar represents the relative energy. The single point energy is calculated with PBE0+vdW, whereas VIP and VEA are determined with G0W0@PBE0.}
	\label{fig0}
\end{figure}

\begin{center}
	{\large \bf Supporting Information}\\ 
\end{center}
\vspace*{12pt}
\noindent {\bf \large I. VEA, VIP and Relative energy of pristine (TiO$_2$)$_{n}$ clusters\\}
\noindent {\bf \large II. Minimum energy isomers of N-doped (TiO$_2$)$_{10}$ cluster\\}
\noindent {\bf \large III. Density of states for (TiO$_2$)$_{10}$ cluster\\}
\noindent {\bf \large IV. 2D phase diagrams for (GA)$_\textrm{E}$, (GA)$_\textrm{P}$$^\textrm{EA}$ and (GA)$_\textrm{P}$$^\textrm{IP}$ based N- doped (TiO$_2$)$_{n}$ clusters, where n = 5, 10, 15, 20 \\}
\noindent {\bf \large V. Excitation energy of undoped  (TiO$_2$)$_{n}$ clusters\\}
\noindent {\bf \large VI. TDOS and PDOS for (GA)$_\textrm{E}$ and (GA)$_\textrm{P}$$^\textrm{IP}$ based (un)doped  (TiO$_2$)$_{10}$ clusters\\}
%\clearpage
%\newpage
%%%%%%%%%%%%%%
\newpage
\noindent {\bf \large I. VEA, VIP and Relative energy of pristine (TiO$_2$)$_{n}$ clusters\\}
In Fig~\ref{fig0}(a, b, c), we have plotted VEA, VIP and relative energy of undoped clusters which are scanned by (GA)$_\textrm{E}$, (GA)$_\textrm{P}$$^\textrm{EA}$ and (GA)$_\textrm{P}$$^\textrm{IP}$ algorithm, respectively.
Each GA has scanned the different regions of the configuration space. The energy based algorithm [(GA)$_\textrm{E}$] has scanned the clusters with minimum energy (see Fig~\ref{fig0}a), whereas the property based algorithms [(GA)$_\textrm{P}$$^\textrm{EA}$ and (GA)$_\textrm{P}$$^\textrm{IP}$] have scanned the clusters with higher energy than that obtained by (GA)$_\textrm{E}$ as shown in Fig~\ref{fig0}b and ~\ref{fig0}c. 

\noindent {\bf \large II. Minimum energy isomers of N-doped (TiO$_2$)$_{10}$ cluster\\}
\begin{figure}[h!]
	\includegraphics[width=1.0\columnwidth,clip]{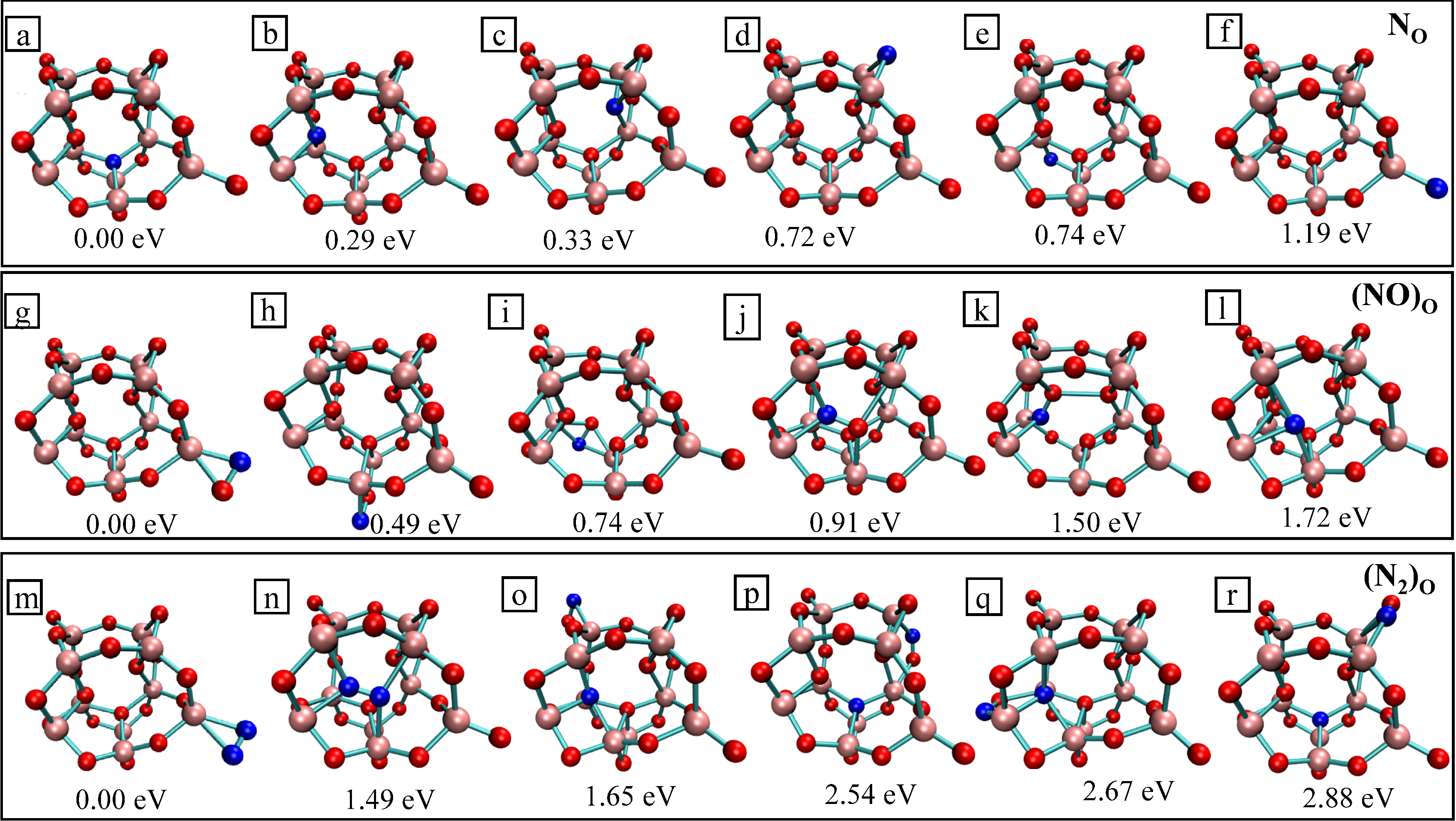}
	\caption{(color online) Relative energy in eV with respect to minimum energy isomers of the respective N-doped (TiO$_2$)$_{10}$ clusters: (a-f) N$_\textrm{O}$, (g-l) (NO)$_\textrm{O}$ and (m-r) (N$_2)_\textrm{O}$.}
	\label{fig1}
\end{figure}
In Fig~\ref{fig1}, we have shown some structures of different types of N-doped (TiO$_2$)$_{10}$ clusters and their relative DFT energies with respect to minimum energy structure. In the upper, middle and lower panel, different configurations of N$_\textrm{O}$, (NO)$_\textrm{O}$, and (N$_2)_\textrm{O}$ are shown, respectively.\\

\newpage
\noindent {\bf \large III. Density of states for undoped (TiO$_2$)$_{10}$ cluster\\}
\begin{figure}[h!]
	\includegraphics[width=0.5\columnwidth,clip]{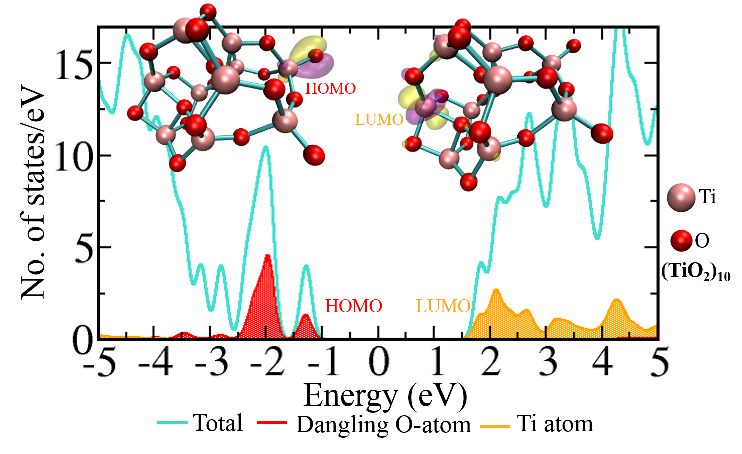}
	\caption{TDOS and PDOS for (TiO$_2$)$_{10}$ cluster. Localized electronic density of HOMO and LUMO states are shown here.}
	\label{fig2}
\end{figure}
In Fig~\ref{fig2}, we have shown the TDOS, and PDOS of the atoms, which are mainly contributing at HOMO and LUMO level. At HOMO level, states are highly localized to the dangling O-atoms, whereas LUMO states are localized to the Ti-atom, which is at the maximum distance from the dangling O.\\
  
\begin{figure}[h!]
	\includegraphics[width=1.0\columnwidth,clip]{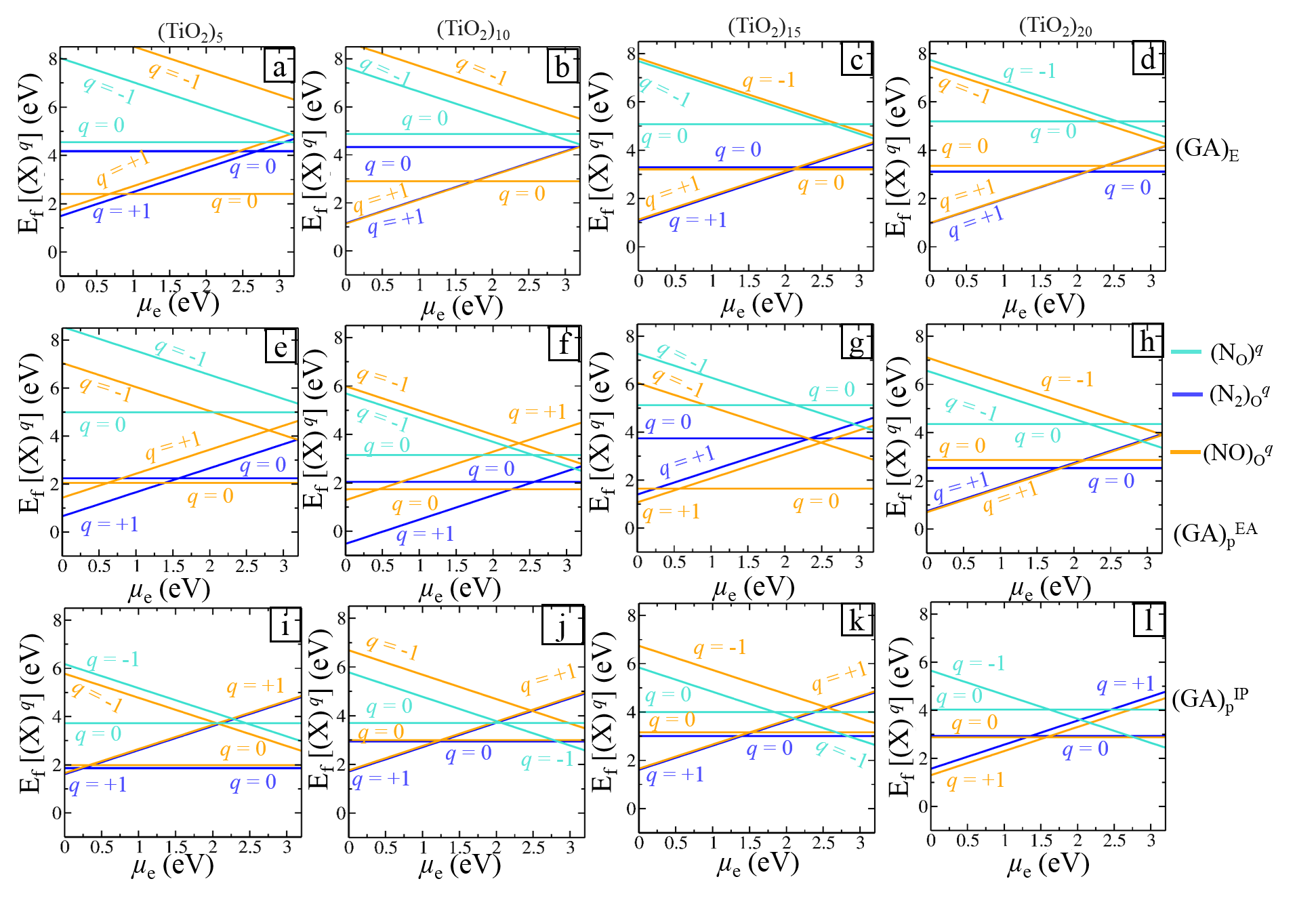}
	\caption{2D phase diagrams of N-doped (TiO$_2$)$_{10}$ clusters at $T$ = 300 K and $p_{{\textrm O}_2}$ = 1 atm. The formation free energy is shown as a function of chemical potential of electron ($\mu_\textrm{e}$). The upper, middle and lower panel of the phase diagrams are for (GA)$_\textrm{E}$, (GA)$_\textrm{P}$$^\textrm{EA}$ and (GA)$_\textrm{P}$$^\textrm{IP}$ based clusters, respectively, for size: (a, e, i) n = 5, (b, f, j) n = 10, (c, g, k) n = 15, (d, h, l) n = 20.}
	\label{fig3}
\end{figure}

\noindent {\bf \large IV. 2D phase diagrams for (GA)$_\textrm{E}$, (GA)$_\textrm{P}$$^\textrm{EA}$ and (GA)$_\textrm{P}$$^\textrm{IP}$ based N- doped (TiO$_2$)$_{n}$ clusters, where n = 5, 10, 15, 20\\}
In Fig~\ref{fig3}, we have shown the 2D phase diagrams of N-doped (TiO$_2$)$_{10}$ clusters at ambient condition ($T$ = 300 K and $p_{{\textrm O}_2}$ = 1 atm). In the upper, middle and lower panel, phase diagrams are shown for (GA)$_\textrm{E}$, (GA)$_\textrm{P}$$^\textrm{EA}$ and (GA)$_\textrm{P}$$^\textrm{IP}$ based clusters, respectively. In (GA)$_\textrm{E}$ clusters for n = 5, at lower values of $\mu_\textrm{e}$ (p-type doping region), (N$_2$)$_\textrm{O}^{+1}$ is the most stable state with the minimum formation energy, whereas at higher values of $\mu_\textrm{e}$ (n-type doping region) (NO)$_\textrm{O}^{0}$ phase becomes stable as shown in Fig~\ref{fig3}a. For n = 10, 15 and 20, (N$_2$)$_\textrm{O}^{+1}$ and (NO)$_\textrm{O}^{+1}$ states have the same formation energy near p-type doping. Near n-type doping region, (NO)$_\textrm{O}^{0}$ is the most stable for n = 10 and 15, whereas (N$_2$)$_\textrm{O}^{0}$ is stable for n = 20 (see the Fig~\ref{fig3}(b-d)). In (GA)$_\textrm{P}$$^\textrm{EA}$ case, near p-type doping region (N$_2$)$_\textrm{O}^{+1}$ is the most sable phase for n = 5, 10, and near n-type doping region (NO)$_\textrm{O}^{0}$ becomes favorable as shown in Fig~\ref{fig3}(e-f). For n = 15, (NO)$_\textrm{O}^{+1}$ and (NO)$_\textrm{O}^{0}$ are the stable phases in p-type and n-type doping region, respectively. Note that, for n = 20, (N$_2$)$_\textrm{O}^{+1}$ and (NO)$_\textrm{O}^{+1}$ phases are overlapping in p-type doping region. Near n-type doping region, (N$_2$)$_\textrm{O}^{0}$ becomes the most stable. In N-doped (GA)$_\textrm{P}$$^\textrm{IP}$ clusters, (NO)$_\textrm{O}^{+1}$ and (N$_2$)$_\textrm{O}^{+1}$ both phases have nearly same formation energy and are stable in p-type doping region for size n = 5, 10 and 15. However, for size n = 20, (NO)$_\textrm{O}^{+1}$ is the most stable phase in p-type doping region. In n-type doping region, for n = 5, (N$_2$)$_\textrm{O}^{0}$ phase is most stable. For size = 10, 15 and 20, N$_\textrm{O}^{-1}$ state is the most stable state having the minimum formation energy.
\newpage
\noindent {\bf \large V. Excitation energy of undoped  (TiO$_2$)$_{n}$ clusters\\}
\begin{table}[h!]
	% 	\centering
	\begin{tabular}{ |c|c|c|c|c|c|c| } 			 
		\hline	
		\begin{large}Cluster size\end{large}
		& \begin{small}E$_x$ [(GA)$_\textrm{E}$]\end{small} & \begin{small}E$_x$ [(GA)$_\textrm{P}$$^\textrm{EA}$]\end{small} & \begin{small}E$_x$ [(GA)$_\textrm{P}$$^\textrm{IP}$]\end{small} & \begin{small}E$_{exp}$\cite{zhai2007probing} \end{small}
		& \begin{small}E$_{TD}$~\cite{shevlin2010electronic}\end{small} & \begin{small}E$_{HL}$\cite{qu2006theoretical,qu2007theoretical} \end{small}   \\
		\hline
		(TiO$_{2}$)$_{4}$ &4.33&2.98& 2.60 & 2.60 & 3.30 & 3.15  \\
		\hline 
		(TiO$_{2}$)$_{5}$ &4.19&2.51&2.82 & 2.85 & 3.06 & 4.54 \\
		\hline
		(TiO$_{2}$)$_{6}$&4.23& 2.65 & 2.18 & 3.00 & 3.23 & 4.53  \\
		\hline
		(TiO$_{2}$)$_{7}$&4.65& 3.07 & 2.52 & 3.10 & 2.88 & 3.96  \\
		\hline
		(TiO$_{2}$)$_{8}$&4.59& 3.59 & 2.50 & 3.13 & - & 3.58  \\ 
		\hline
		(TiO$_{2}$)$_{9}$&4.30& 3.59 &  3.35  & 3.13 & - & 3.84  \\
		\hline 
		(TiO$_{2}$)$_{10}$&4.54& 2.73 & 3.15 & 3.31 & 3.14 & 4.49  \\

		\hline
	\end{tabular}  
	\caption{Theoretically calculated excitation energy (E$_x$), experimental excitation energy (E$_{exp}$), singlet transition energy (E$_{TD}$),  and HOMO-LUMO energy E$_{HL}$ corresponding to (TiO$_2$)$_n$ clusters. All the values are in eVs.} 
\end{table}
 
\noindent {\bf \large VI. TDOS and PDOS for (GA)$_\textrm{E}$ and (GA)$_\textrm{P}$$^\textrm{IP}$ based (un)doped  (TiO$_2$)$_{10}$ clusters\\}
\begin{figure}[h!]
	\includegraphics[width=1.0\columnwidth,clip]{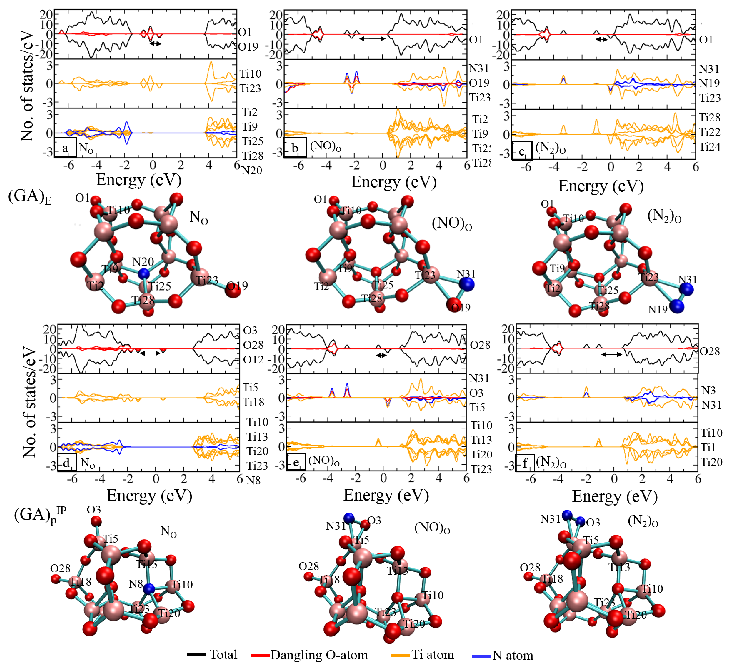}
	\caption{TDOS and PDOS for (GA)$_\textrm{E}$ and (GA)$_\textrm{P}$$^\textrm{IP}$ based (TiO$_2$)$_{10}$ clusters for doped cases: (a, d) N$_\textrm{O}$, (b, e) (NO)$_\textrm{O}$, (c, f) (N$_2)_\textrm{O}$. The respective electronic configurations of doped clusters are shown below the figures. Double headed arrows are depicting the HOMO-LUMO gap}
	\label{fig4}
\end{figure}
In Fig~\ref{fig4}, we have shown the TDOS and PDOS for N-doped (TiO$_2$)$_{10}$ clusters based on (GA)$_\textrm{E}$ and (GA)$_\textrm{P}$$^\textrm{IP}$. TDOS and PDOS of dangling O-atoms are shown in the upper panel. The PDOS of N and Ti-atoms mainly contribute near the HOMO-LUMO levels as visible in the middle and lower panel. In case of (N)$_\textrm{O}$, N-atom is coordinated with 4 Ti-atoms via strong bond as shown in the Fig~\ref{fig4}a and~\ref{fig4}d. Here, dopant act as a deep donor as N-states are away from the Fermi level (see the lower panel of Fig~\ref{fig4}a and~\ref{fig4}d). Consequently, the unoccupied mid gap/trap states are introduced by dangling O-atoms, which degrade the HER potential of the clusters and limits their application in water splitting. For (NO)$_\textrm{O}$ doping in Fig~\ref{fig4}b, we observe the localized states of N-dopants overlapped with bonded Ti and O-atoms above the HOMO level. As a result, we notice the low fundamental gap for this case and get the suitable potential for HER and OER.  However, in Fig~\ref{fig4}c,~\ref{fig4}e and~\ref{fig4}f, we observe occupied deep mid gap states associated with Ti-atoms due to which they are not suitable for OER. 
\newpage
\bibliography{references2}